\documentclass[a4paper,11pt]{article}

\usepackage{amsmath}
\usepackage{amssymb}

\usepackage[colorlinks=true, allcolors=blue]{hyperref}
\usepackage{graphicx}
\usepackage{dcolumn}
\usepackage{bm}

\usepackage[dvipsnames]{xcolor}

\newcommand{\bc}[1]{\textcolor{black}{#1}}

\newcommand{\dg}[1]{${#1}^{\circ}$}
\usepackage{fancyhdr}
\usepackage[titletoc]{appendix}
\usepackage{multicol}
\usepackage{subcaption}
 
\usepackage{bm}
\usepackage{amsfonts}
\usepackage{epstopdf}
\usepackage{xcolor}
\usepackage{color}
\usepackage{wasysym}
\usepackage{hhline}
\usepackage[capitalise]{cleveref}
\usepackage{authblk}
\usepackage{upgreek}
\newcommand{\micron}{{\upmu\mathrm{m}}}

\usepackage{makecell}

\begin{document}

\title{Electron pulse train accelerated by a linearly polarized Laguerre-Gaussian laser beam}


\author[1,2]{Yin Shi}
\author[2]{David R Blackman}
\author[3]{Ping Zhu}
\author[2]{Alexey Arefiev}%
\affil[1]{Department of Plasma Physics and Fusion Engineering, University of Science and Technology of China, Hefei 230026, China}%
\affil[2]{Department of Mechanical and Aerospace Engineering, University of California at San Diego, La Jolla, CA 92093, USA}%
\affil[3]{National Laboratory on High Power Laser and Physics, Shanghai Institute of Optics and Fine Mechanics, Chinese Academy of Sciences, Shanghai 201800, China}

\date{\today}
\vskip -2.0cm
\maketitle
\vskip -3.0cm

\begin{abstract}
A linearly polarized Laguerre-Gaussian (LP-LG) laser beam with a twist index $l = -1$ has field structure that fundamentally differs from the field structure of a conventional linearly polarized Gaussian beam. Close to the axis of the LP-LG beam, the longitudinal electric and magnetic fields dominate over the transverse components. This structure offers an attractive opportunity to accelerate electrons in vacuum. It is shown, using three dimensional particle-in-cell simulations, that this scenario can be realized by reflecting an LP-LG laser off a plasma with a sharp density gradient. The simulations indicate that a 600~TW LP-LG laser beam effectively injects electrons into the beam during the reflection. The electrons that are injected close to the laser axis experience a prolonged longitudinal acceleration by the longitudinal laser electric field. The electrons form distinct monoenergetic bunches with a small divergence angle. The energy in the most energetic bunch is 0.29~GeV. The bunch charge is 6~pC and its duration is $\sim 270$~as. The divergence angle is just \dg{0.57} (10~mrad). By using a linearly polarized rather than a circularly polarized Laguerre-Gausian beam, our scheme makes it easier to demonstrate the electron acceleration experimentally at a high-power laser facility.

\end{abstract}


\section{Introduction} \label{Sec-1}

The construction of numerous high-power laser systems around the world~\cite{Danson2019, elinp10pw2022, Shen2018, Li2022} has enabled the development of novel particle~(see \cite{Bulanov_POP_2016, Esarey2009} and refs. therein) and radiation sources~\cite{Nakamura_PRL_2012, Ridgers_PRL_2012, Ji2014, Stark_PRL_2016, Tao2020, Capdessus2018} for multidisciplinary applications~\cite{Bulanov_2014, Weeks_1997}. Most of the improvements of the laser beams used for driving laser-matter interactions have been focused on increasing power, on-target intensity, total energy, and the contrast of the compressed pulse. For example, the proposed facility~\cite{Shen2018,Li2022} that aims to cross the 100~PW limit is expected to be in development over the next decade.  Concurrently, new optical techniques for producing helical wave-fronts~\cite{Leblanc2017, shi2014,Longman2017, Longman2020,Liang2020,Elkana2022} are also being developed. There now exist multiple computational~\cite{vieira2016, Zhang2015, vieira2018, Shi2018, Longman2017, Ju2018, Zhu2019, TIKHONCHUK2020, Nuter2018, Blackman2020} and experimental~\cite{Leblanc2017,Longman2020,Denoeud2017, BAE2020, Aboushelbaya2020} studies examining interactions of helical laser beams with plasmas. There are also some published works on the terawatt scale helical laser production using a chirped-pulse amplification system~\cite{pan2020, chen2022}. Even though the techniques for creating helical beams with higher power are yet to be applied at high-power high-intensity laser facilities, they offer an exciting opportunity to create laser pulses with a qualitatively different field typology that can have a profound impact on laser-plasma interactions and particle acceleration.

There are several laser-based electron acceleration approaches with a different degree of maturity. The most frequently used ones are the laser wakefield acceleration~\cite{Esarey2009} that utilizes plasma electric fields and the direct laser acceleration~\cite{gibbon2004short} that relies on the fields of the laser for the energy transfer inside a plasma (e.g. see \cite{Arefiev_POP_2016}) or in vacuum~\cite{Stupakov_PRL_2001}. These mechanism are typically realized using conventional laser pulses. In an attempt to improve electron acceleration, several studies also considered radially polarized laser beams~\cite{Zaim2017} and higher-order Gaussian beams~\cite{Sprangle1996}. Recently, there has been an increased interest in utilizing ultra-high-intensity laser beams with helical wave-fronts for electron acceleration in various setups, including vacuum acceleration~\cite{Shi2021, shi2021a, Blackman2022, Pae2022}, {laser wakefield acceleration~\cite{vieira2014, zhang_gb2016}, and microstructural target electron acceleration~\cite{Hu2018b}.}

Conventional high-power laser systems~\cite{Danson2019} generate linearly polarized (LP) laser beams without a twist to the laser wave-fronts, which prevents one from readily realizing the interactions utilizing ultra-high-intensity helical laser beams. The spatial structure of laser beams with helical wave-fronts can be viewed as a superposition of Laguerre-Gaussian modes, which is why these beams are often referred to as Laguerre-Gaussian or LG beams. An LG beam can potentially be produced from a standard LP Gaussian laser pulse in reflection from a fan-like structure~\cite{shi2014,Longman2017,Elkana2022}. This approach avoids transmissive optics and it is well-suited for generating high-power high-intensity LG beams at high efficiency. Achieving circularly polarized (CP) LG beams at conventional high-intensity laser systems is likely to be more challenging than achieving LP-LG beams since a native Gaussian beam is linearly polarized and extra steps need to be taken to induce circular polarization. It is then imperative to study laser-plasma interactions involving LP-LG beams, as these beams are more likely to be achieved in the near-term at high-intensity laser systems like ELI-NP~\cite{elinp10pw2022} or the SG-II UP facility~\cite{zhu2018hpl}. 

The focus of this paper is on electron acceleration in vacuum by an LP-LG laser beam following its reflection off a plasma with a sharp density gradient.
{This setup is sometimes referred to as the ``reflection off a plasma mirror''~\cite{Thevenet2016}, but we minimize the use of the term ``plasma mirror'' to avoid any confusion with optical shutters employed for producing high contrast pulses.} Direct laser acceleration in vacuum by a conventional laser beam is generally considered to be ineffective. The key issue is the transverse electron expulsion caused by the transverse electric field of the laser. The expulsion terminates electron energy gain from the laser and leads to strong electron divergence. It must be stressed that the expulsion is closely tied to the topology of the laser field that is dominated by the transverse electric and magnetic fields. In two recent publications~\cite{Shi2021, shi2021a} we showed that a CP-LG beam with a properly chosen twist can be used to solve the expulsion problem. The wave front twist creates a unique accelerating structure dominated by longitudinal laser electric and magnetic fields in the region close to the axis of the beam. The longitudinal electric field provides forward acceleration without causing electron divergence, while the longitudinal magnetic field provides transverse electron confinement. It was shown using 3D particle-in-cell (PIC) simulations that a CP-LG beam reflected off a plasma can generate dense bunches of ultra-relativistic electrons via the described mechanism~\cite{Shi2021, shi2021a}. 
\bc{The distinctive features of this acceleration mechanism are the formation of multiple sub-$\micron$ electron bunches, their relatively short acceleration distance (around $100~\micron$), and their high density (in the range of the critical density).} The purpose of the current study is to identify the changes introduced by the change in polarization from circular to linear with the ultimate goal of determining whether the use of circular polarization is essential.

In this paper, we present results of a 3D PIC simulation for a 600~TW LP-LG laser beam reflected off a plasma with a sharp density gradient. 
We find that, despite the loss of axial symmetry introduced by switching from circular to linear polarization, the key features of electron acceleration are retained. Namely, the laser is still able to generate dense ultra-relativistic electron bunches, with the acceleration performed by the longitudinal laser electric field in the region close to the laser axis. In the most energetic bunch, the electron energy reaches 0.29~GeV (10\% energy spread). The bunch has a charge of 6~pC, a duration of $\sim 270$~as, and remarkably low divergence of \dg{0.57} (10~mrad).  The normalized emittance in $y$ and $z$ is $\widetilde{\epsilon}_{rms, y} \approx 5 \times 10^{-7}, \widetilde{\epsilon}_{rms, z} \approx 4 \times 10^{-7}$. 

\bc{Such dense attosecond bunches can find applications in research and technology~\cite{Norbert2019, Black2019}, with one specific application being free-electron lasers~\cite{Huang2012}.} The rest of this paper is organized as follows. \Cref{Sec-2} presents the field structure of the LP-LG beam and the setup of our 3D PIC simulation. \Cref{Sec-3} discusses the formation of electron bunches that takes place during laser reflection off the plasma. \Cref{Sec-4} examines the energy gain by the electron bunches during their motion with the laser pulse. \Cref{Sec-5} summarizes our key results and discusses their implication. 

\section{Field structure of the LP-LG beam and simulation setup}\label{Sec-2}

\begin{figure}[!t]
 \centering
 \includegraphics[width=0.85\linewidth]{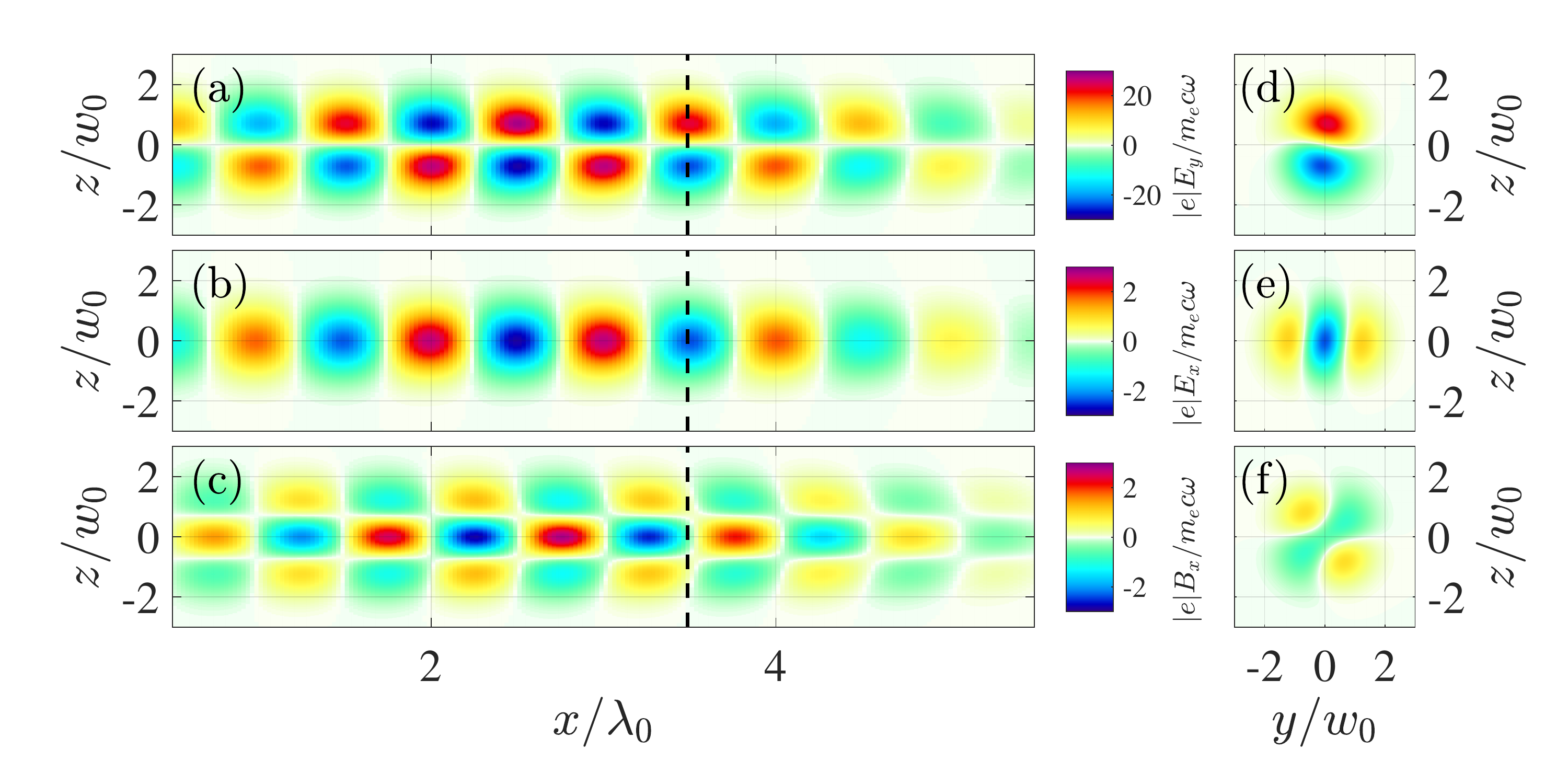} 
\caption{Electric and magnetic field components of an LP-LG laser beam {before} it encounters the plasma. 
Panels (a) and (d) show $E_y$; panels (b) and (e) show $E_x$; panels (c) and (f) show $B_x$. The left column [(a), (b), and (c)] shows the field structure in the $(x,z)$-plane at $y = 0$. The right column [(d), (e), and (f)] shows the field structure in the $(y,z)$-plane at the $x$-position indicated with the dashed line in panels (a) - (c). All the snapshots are taken at $t \approx -9$~fs from the simulation with parameters listed in \cref{table:PIC}. } \label{eb_xz}
\end{figure}

In this section, we present the structure of the LP-LG beam that we use in our 3D PIC simulation to generate and accelerate electron bunches. The section also presents the simulation setup.

The wave front structure of a helical beam can be parameterized using two indices: the twist index $l$ that specifies the azimuthal dependence of the {transverse} electric and magnetic fields and the radial index $p$ that specifies the radial dependence of the same fields in the focal plane. The polarization of the transverse laser fields is independent of their wave front topology, so a helical beam can be linearly or circularly polarized. Detailed expressions for all field components of an LP-LG beam are provided in Ref.~\cite{shi2021a}. We choose to omit these expressions here for compactness and instead we summarize the key features. The twist index $l$ qualitatively changes the topology of the transverse and longitudinal fields. We are interested in the field structure close to the central axis. There are three distinct cases: $l = 0$, $|l| = 1$, and $|l| > 1$. The case of $l = 0$ corresponds to a conventional beam, with the near-axis field structure dominated by transverse electric and magnetic fields. In the case of $|l| > 1$, all laser fields vanish on the central axis. The case that is of interest to us 
is the case with $|l| = 1$, because in this case the longitudinal rather than transverse fields peak on axis.

In our 3D PIC simulations, we use a linearly polarized 600~TW laser with $l = -1$ and $p = 0$. We consider a beam that propagates in the negative direction along the $x$-axis upon entering the simulation box. Detailed parameters of the laser beam are listed in \cref{table:PIC}. In order to facilitate a comparison with the results for a CP-LG beam published in Ref.~\cite{shi2021a} (right circular polarized with $l = -1$ and $p=0$), we use the same peak power, pulse duration, and spot size for our LP-LG beam. The electric and magnetic field structure of the LP-LG beam in the $(x,z)$-plane and the $(y,z)$-plane are shown in~\cref{eb_xz}. The plots illustrate the difference in topology between the transverse and longitudinal field components. The longitudinal electric and magnetic fields reach their highest amplitude along the axis of the laser beam [see \cref{eb_xz}(b) and \cref{eb_xz}(c)]. On the other hand, the transverse electric field shown in \cref{eb_xz}(a) vanishes on the axis of the beam. The electric field structure in \cref{eb_xz} agrees with the analytical expression given in~\cref{append:theory} and derived in paraxial approximation by assuming that the diffraction angle $\theta_d$ is small. The longitudinal fields are relatively strong even though $\theta_d \ll 1$. We have $E_{\parallel}^{\max} /E_{\perp}^{\max} = B_{\parallel}^{\max} /B_{\perp}^{\max} \approx 0.14$ for the considered LP-LG beam with $\theta_d \approx 8.5 \times 10^{-2}$, where $\theta_d = \lambda_0 / \pi w_0$. Note that $E_{\parallel}^{\max} \approx 1.1 \times 10^{13}$~V/m and $B_{\parallel}^{\max} \approx 36$~kT. The peak normalized amplitude of the longitudinal field $a_* = |e| E_{\parallel}^{\max} / m_e c \omega$ for a given period-averaged power $P$ in PW is given by \cref{a of P}, where $c$ is the speed of light, $\omega = 2 \pi c / \lambda_0$ is the laser frequency, and $e$ and $m_e$ are the electron charge and mass. We find that $a_* \approx 2.7$ for the considered power of 600~TW. 

It is instructive to compare the field structure of the LP-LG beam to the field structure of the CP-LG beam from Ref.~\cite{shi2021a}. In both cases, longitudinal electric and magnetic fields peak on the axis of the beam where transverse field vanish. However, in contrast to the CP-LG beam, the longitudinal electric and magnetic fields of the LP-LG beam lack axial symmetry [see \cref{eb_xz}(e) and \cref{eb_xz}(f)]. The difference in symmetry can be illustrated by constructing an LP-LG beam ($l = -1$) from two co-propagating CP-LG beams. Following the notations of Ref.~\cite{shi2021a}, we create a circularly-polarized transverse electric field by adding $E_z = i \sigma E_y$, where $\sigma = 1$ produces a right-circularly polarized wave and $\sigma = -1$ produces a left-circularly polarized wave. We take a pair of CP-LG beams: one with $l = -1$, $\sigma = -1$ and another one with $l = -1$, $\sigma = 1$. Their superposition produces a linearly polarized transverse electric field, because $E_z$ components of these beams cancel each other out. It was shown in Refs.~\cite{Shi2021} and \cite{shi2021a} that the longitudinal fields of the right and left CP-LG beams have different dependencies on $r$ and $\phi$. Specifically, the longitudinal field of the right CP-LG beam ($l \sigma = -1$) is axisymmetric, whereas the longitudinal field of the left CP-LG beam ($l \sigma = 1$) has azimuthal dependence. Moreover, only the right CP-LG beam contributes to the longitudinal fields on the axis, because the fields of the left CP-LG beam vanish. The LP-LG beam inherits its azimuthal dependence from the left CP-LG beam, which is the reason why the longitudinal fields still peak on the axis, but lose their symmetry as we move away from the axis. 

As stated earlier, we want to contrast our results with those for a right CP-LG beam that has the same power. The key player in electron acceleration is the longitudinal electric field of the laser, because it is this field that performs most of the acceleration for the electrons moving along the laser axis. We use the discussed decomposition for the LP-LG beam to compare the longitudinal field strength $|E_x|$ on the axis of the two beams. We have already shown that $|E_x|$ of an LP-LG beam is equal to $|E_x|$ of a right CP-LG beam whose transverse field amplitude is half of that in the LP-LG beam. The power of the right CP-LG beam is two times lower than the power the LP-LG beam. Since the power scales as the square of the field strength, we immediately conclude that $|E_x|$ in a right CP-LG beam whose power is the same as the power of the LP-LG beam is going to be higher by a factor of $\sqrt{2}$. The loss of axial symmetry and the reduced field strength are likely to alter the injection and subsequent acceleration of electron bunches by the LP-LG beam compared to the case of the right CP-LG beam from Ref.~\cite{shi2021a}.

In our simulation performed using PIC code EPOCH~\cite{Arber2015}, the discussed LP-LG beam is reflected off a plasma with a sharp density gradient. In what follows, we provide details of the simulation setup that we use in the next sections to study electron injection and acceleration. The target is initially set as a fully ionized carbon plasma with electron density $n_e = 180 n_c$, where $n_c$ = 1.8$\times 10^{27}$~m$^{-3}$ is the critical density for the considered laser wavelength $\lambda_0 = 0.8~\mathrm{\upmu m}$. \Cref{table:PIC} provides details regarding the density gradient. The focal plane of the beam in the absence of the plasma is located at $x = 0~\mathrm{\upmu m}$, which is also the location of the plasma surface. The resolution and the number of particles per cell in the PIC simulation is determined based on a convergence test discussed in \cref{append:conv}. The test addresses a concern that the parameters of the accelerated electron bunches might be sensitive to simulation parameters~\cite{Shi2021}.

\begin{table}
\centering
\begin{tabular}{ |m{7cm}|m{6.7cm}| }
 \hline
 \multicolumn{2}{|l|}{\textbf{Parameters for linearly polarized Laguerre-Gaussian laser} }\\
 \hline
 Peak power (period averaged) & 0.6~PW \\
 Radial and twist index & $p = 0, l = -1$ \\
 Wavelength & $\lambda_0 = 0.8$ $\mathrm{\upmu m}$\\
 Pulse duration ($\sin^2$ electric field) & $\tau_g=20$~fs\\
 Focal spot size ($1/\mathrm{e}$ electric field) & $w_0$ = 3~$\mathrm{\upmu m}$\\
 Location of the focal plane & $x = 0~\upmu$m\\
 Laser propagation direction & $-x$ \\
 Polarization direction & $y$ \\
 \hline \hline
 \multicolumn{2}{|l|}{\textbf{Other simulation parameters} }\\
 \hline
 Position of the foil and the pre-plasma & $-1.0 \sim -0.3~\mathrm{\upmu m} $ and $-0.3\sim$0.0~$\mathrm{\upmu m}$\\
 The density distribution of pre-plasma & $n_e = 180.0 n_c \exp[ - 20 (x + 0.3~\mathrm{\upmu m})/ \lambda_0]$ \\
 Electron and ion (C$^{+6}$) density in foil & $n_e = 180.0 n_{c}$ and $n_i = 30.0 n_{c}$ \\
 Gradient length & $L = \lambda_0$/20\\
 Simulation box ($x \times y \times z$) & 10~$\mathrm{\upmu m}$ $\times$ 20~$\mathrm{\upmu m}$ $\times 20~\mathrm{\upmu m}$\\ 
 Cell number($x \times y \times z$) & 800 cells $\times$ 1600 cells $\times$ 1600 cells \\
 Macroparticles per cell for electrons & 100 at $r <$ 2.5~$\mathrm{\upmu m}$, 18 at $r \geq 2.5~\mathrm{\upmu m}$ \\
 Macroparticles per cell for C$^{+6}$ & 12 \\
Order of EM-field solver & 4 \\
 \hline 
 \end{tabular}
 \caption{3D PIC simulation parameters. $n_{c} = 1.8\times 10^{27}$~m$^{-3}$ is the critical density corresponding to the laser wavelength $\lambda_0$. The initial temperatures for electrons and ions are set to zero.}
 \label{table:PIC}
\end{table}

\begin{figure}[htb]
 \centering  
 \includegraphics[width=0.7\linewidth]{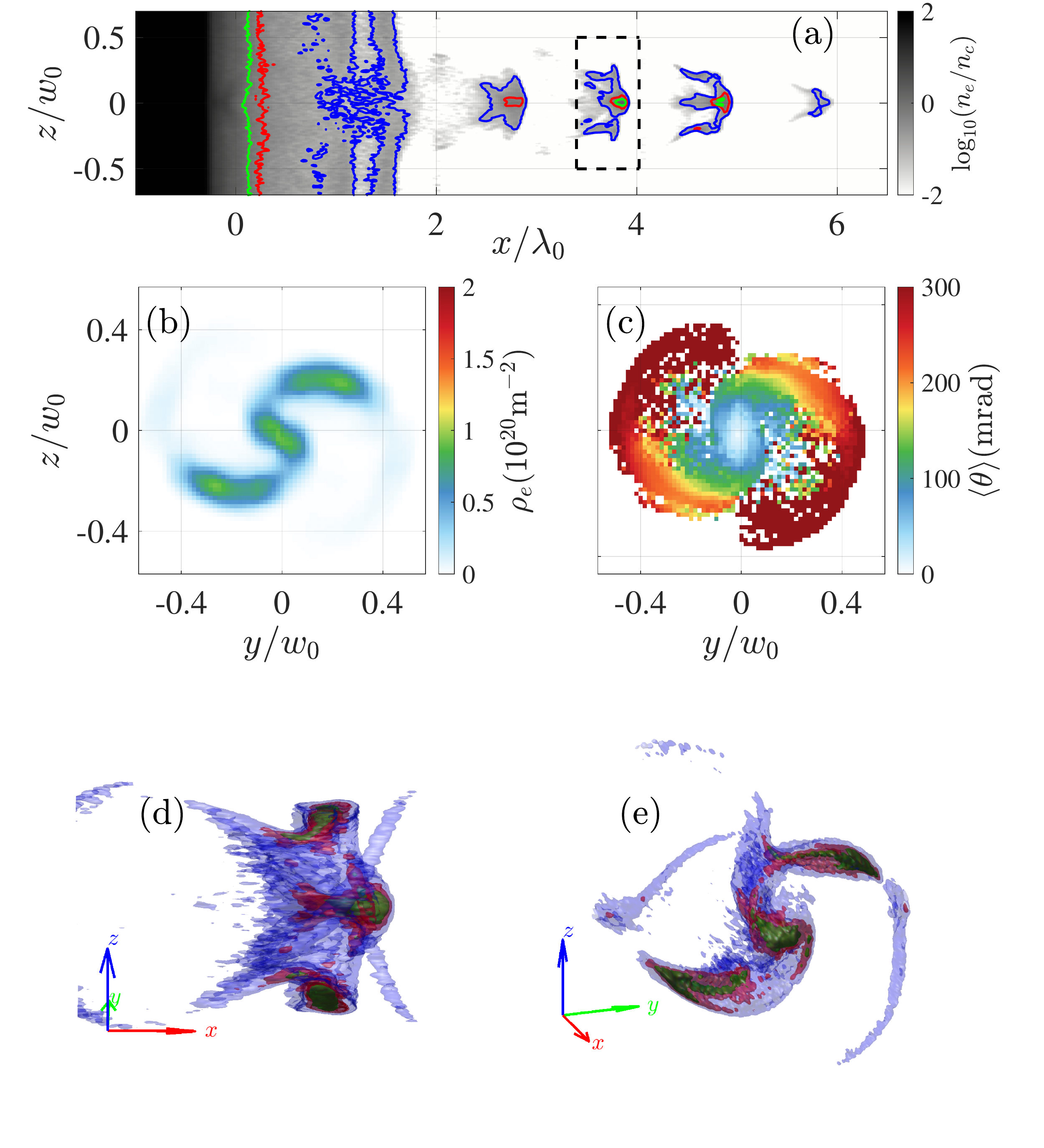} 
\caption{Structure of electron bunches shortly after laser reflection off the plasma ($t = 9$~fs). (a)~Electron density on a log-scale, with the color representing $\log(n_e/n_c)$. The blue, red, and green contours denote $n_e = 0.1n_{c}$, $0.5 n_{c}$, and $n_{c}$. The dashed rectangle marks the third bunch whose additional details are provided in the remaining panels. 
(b) Electron areal density $\rho_e$ in the third bunch. (c) Cell-averaged electron divergence angle $\langle \theta \rangle$ in the third bunch.  (d)\&(e) 3D rendering of the electron density in the third bunch using different viewpoints.} \label{edenst9fs}
\end{figure}


\section{Electron injection into the LP-LG laser beam}\label{Sec-3}

In this section we discuss the formation of electron bunches that takes place during laser reflection off the plasma. We refer to this process as the `electron injection', because, once the bunches are formed, they continue surfing with the laser beam.

\Cref{edenst9fs} shows various aspects of electron injection. All snapshots are taken at $t = 9$~fs, with $t = 0$~fs being defined as the time when the peak of the laser envelope reaches $x = 0$ (in the absence of the plasma). The electron density, $n_e$, in the $(x,z)$-plane is shown in \cref{edenst9fs}(a). At this point, most of the laser beam (incident from the right) has been reflected by the plasma. The reflection process generates bunches that are solid in the near-axis region, with the peak densities as high as $n_c$. {The plot of $n_e$ integrated over the laser beam cross section that is shown in \cref{Comp9fs261fs}(a) of \cref{append:conv} provides additional information about the bunches.} \Cref{edenst9fs}(b) and \cref{edenst9fs}(c) show the transverse areal density  $\rho_e$ and cell-averaged divergence angle $\langle \theta \rangle$ of the third bunch marked with a dashed rectangle in \cref{edenst9fs}(a). The divergence angle for an individual electron is defined as $\theta \equiv \arctan(p_{\bot}/p_{x})$. The angle is averaged on every mesh cell of the $(y,z)$-plane. It is instructive to compare the plots of $\rho_e$ and $\langle \theta \rangle$ to the results for the CP-LG laser beam presented in Ref.~\cite{shi2021a} for the same time instant ($t = 9$~fs). In the case of the LP-LG beam, the areal density in the near-axis region is approximately two times lower while the divergence angle in the same region is similar to that of the CP-LG beam.
The biggest difference is the loss of axial symmetry. The LP-LG beam generates two dense side-lobes in addition to the on-axis part that has been discussed.
To further illustrate the complex structure of the electron bunches generated by the LP-LG beam, \cref{edenst9fs}(d) and \cref{edenst9fs}(e) provide 3D rendering of electron density in the third bunch. 
A corresponding movie with different viewpoints can be found in the Supplemental Material. It is clear from \cref{edenst9fs}(d) that the two side-lobes are slightly behind the on-axis region, which means that the phase of the laser field at their location is different.
The divergence of the two lobes [shown in \cref{edenst9fs}(c)] is so high that they are likely to move away from the central axis while moving in the positive direction along the $x$-axis with the laser beam. 

\begin{figure}[htb]
 \centering
 \includegraphics[width=0.7\linewidth]{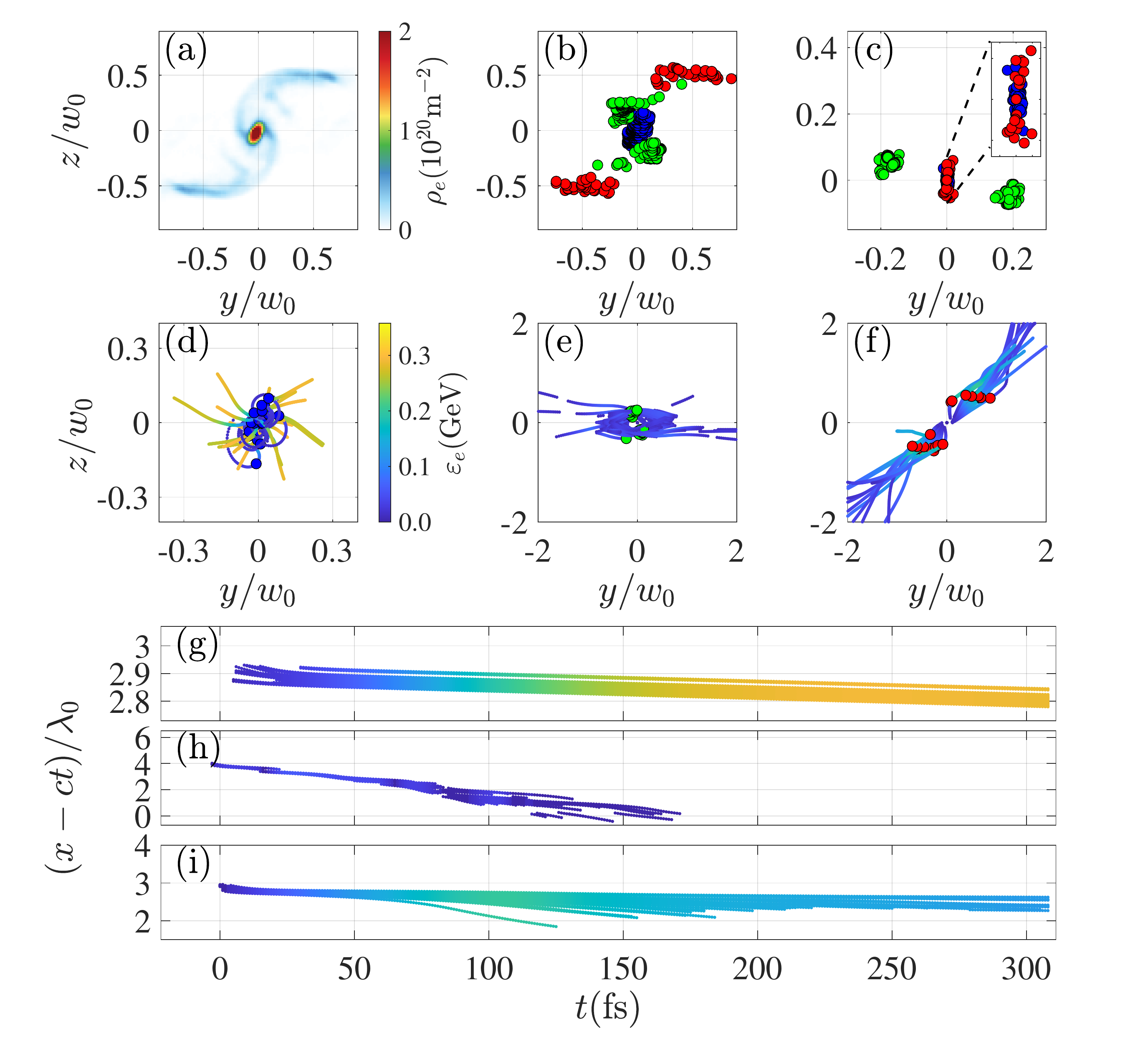}
 \caption{(a) Areal density of the electrons in the third bunch at time  $t = 46$ fs. (b) Three groups of electrons (blue, green and red markers) selected from the third bunch at $t = 46$~fs for tracking. The electrons in each group are selected randomly. (c) Transverse  positions of the three groups of electrons from (b) at $t = -2.2$~fs. (d), (e), \& (f) Trajectories of the three groups of electrons in the transverse plane over the duration of the simulation. The line color shows electron energy. The markers show the electron locations at $t = 46$~fs. (g), (h), \& (i) Time evolution of the longitudinal position for the same three groups of electrons, with (g) showing `blue' electrons, (h) showing `green' electrons, and (i) showing `red' electrons. The line color shows electron energy. }  \label{etraj}
\end{figure}

To examine this expectation and to provide additional insights into electron bunch dynamics, we performed detailed particle tracking for the third bunch. We distinguish three groups of electrons based on their transverse position within the bunch at $t = 46$~fs. \Cref{etraj}(a) shows the areal density of the third bunch at $t = 46$~fs, while \cref{etraj}(b) shows three groups of electrons selected for tracking. The electrons are picked randomly from the entire electron population of the third bunch. Note that we choose $t = 46$~fs rather than $t = 9$~fs as our selection time in order to give enough time for the three groups to become visibly separated. The selected particles are tracked during the entire simulation (up to $t \approx 310$~fs) to determine their trajectories and energy gain. 

Figures~\ref{etraj}(d)-(f) provide projections of electron trajectories onto the beam cross-section, where the color-coding is used to show electron energy along each trajectory. The markers correspond to the electron positions at $t = 46$~fs. To see initial electron positions, we provide \cref{etraj}(c) that shows electron positions in the $(y,z)$-plane at $t = -2.2$~fs. As seen in \cref{etraj}(d), the `blue' electrons remain close to the axis of the laser beam and thus within the region with a strong longitudinal electric field throughout the simulation. The `green' electrons [see \cref{etraj}(e)] rotate around the axis and eventually leave the analysis window [$y \in (-2, 2)w_0$, $z \in (-2, 2)w_0$]. The `red' electrons [see \cref{etraj}(f)] are different because they are expelled directly outwards without any significant rotation. These electrons travel through the region with a significant transverse electric field. 
The long-term energy gain by these three groups of electrons is discussed in the next section.

\section{Electron energy gain in the LP-LG laser beam}\label{Sec-4}


In \cref{Sec-2} we showed that the reflection of an LP-PG beam produces dense electron bunches. These bunches can move with the laser beam gaining energy. In this section we examine this energy gain.

Figures~\ref{etraj}(g)~-~(i) show how the energy of electrons in each group from \cref{etraj}(b) changes over time. The magnitude of the longitudinal electron velocity $v_x$ is a major factor determining the electron energy gain. Electrons with smaller $c - v_x$ can stay longer in the accelerating part of the laser wave front while moving forward with the laser beam. Due to the fact that the considered electrons are ultra-relativistic, it is the divergence angle $\theta$ rather than the magnitude of the velocity that primarily influences $v_x$, with $v_x \approx c \cos \theta$. To assess the difference between $v_x$ and $c$ that can be extremely small, we use the vertical coordinate that shows $(x - ct) / \lambda_0$ in Figs.~\ref{etraj}(g)~-~(i). It is essentially the relative slip (in the units of $\lambda_0$) between the electron and a point moving with the speed of light. The `blue' electrons remain close to the beam axis and have the smallest $c - v_x$. As seen in \cref{etraj}(g), they slip by less than $0.2 \lambda_0$ over 300~fs, which allows them to gain roughly 290~MeV. The `green' electrons have a much bigger value of $c - v_x$ because there is a transverse component of electron velocity associated with the rotation. As seen in \cref{etraj}(h), they experience significant slipping over 100~fs which prevents them from prolonged acceleration required for a substantial energy gain. The `red' electrons have the biggest transverse displacement early on, so that they are exposed to a strong transverse laser electric field. This field causes their transverse motion, but it also transfers energy to the electrons. This is the reason why the `red' electrons shown in \cref{etraj}(i) gain more energy than the `green' electrons. However, their slipping causes them to experience a decelerating field before the laser beam has time to diverge. This is the underlying cause for the energy reduction at $t > 100$~fs. The analysis of electron trajectories leads us to a conclusion that the most energetic electrons in our setup are the electrons that remain close to the axis of the beam. In what follows, we focus on their energy gain.

\begin{figure}
 \centering
 \includegraphics[width=0.9\linewidth]{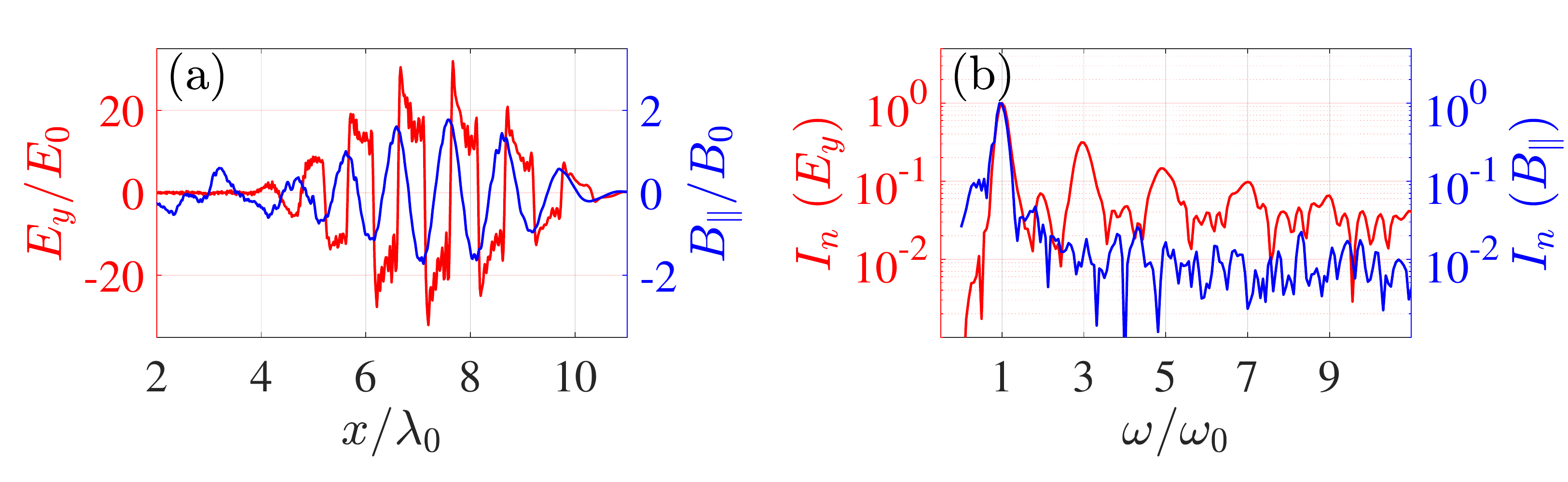} 
\caption{Electric and magnetic fields after reflection of the LP-LG laser beam off the plasma. (a) Longitudinal profiles of the transverse electric field $E_y$ (red curve) and longitudinal magnetic field field $B_{\parallel}$ (blue line) at $t = 21$~fs. $B_{\parallel}$ is plotted along the axis of the beam ($y = 0$, $z = 0$), whereas $E_y$ is plotted at an off-axis location ($y = 0$, $z = 0.7w_0$) where its amplitude has the highest value. (b) Frequency spectra of $E_y$ (red line) and $B_{\parallel}$ (blue line) from panel (a). } \label{EB_hhg}
\end{figure}

Before we proceed with the analysis of the electron acceleration, we take a closer look at the reflected fields. \Cref{EB_hhg}(a) shows transverse electric field away from the axis ($y = 0$, $z = 0.7w_0$) and longitudinal magnetic field on the axis ($y = 0$, $z = 0$) after the laser has been reflected by the plasma. As we have already seen, the electrons tend to bunch on the axis. The field of these electrons is difficult to separate from the longitudinal laser electric field. This is the reason why have plotted $B_{\parallel}$ instead of $E_{\parallel}$. The transverse field is plotted off-axis because it vanishes on the axis of the beam for the considered helical beam. The most striking feature compared to the CP-LG beam examined in~\cite{Shi2021, shi2021a} is the appearance of higher-order harmonics in $E_y$. In contrast to $E_y$, $B_{\parallel}$ has a more regular shape. The spectra shown in \cref{EB_hhg}(b) confirm that $E_y$ contains odd harmonics due to high-harmonic generation effects~\cite{Teubner2009}, whereas $B_{\parallel}$ in the near-axis region seems to be unaffected by the harmonic generation. According to Refs.~\cite{Zhang2015} and \cite{Denoeud2017}, the twist index $l_n$ of the harmonics generated during reflection of an LP-LG laser beam scales as $l_n = n l$, where $n$ is the harmonic order. The field profiles suggest that the analysis of electron acceleration in the near-axis region can be performed without taking into account the higher-order harmonics which have different Gouy phase shift.

The momentum gain of the electrons moving along the axis of the laser beam can then be obtained by integrating the momentum balance equation
\begin{equation}
    d p_{\parallel} / dt = - |e| E_{\parallel},
\end{equation}
where $E_{\parallel}$ is the on-axis component of the laser electric field. We neglect high-harmonic generation and beam scattering, so $E_{\parallel}$ is the real part of the on-axis field in the original beam given by \cref{E longitudinal}. The longitudinal electric field has the same dependence on $x$ as the field of the CP-LG beam considered in Ref.~\cite{shi2021a}. For example, we have
\begin{equation}
    E_{\parallel} = - \frac{E_* \sin (\Phi + \Phi_0)}{1 + x^2 / x_R^2}
\end{equation}
for an electron that is staying close to the peak of the envelope, where $E_*$ is the peak amplitude of $E_{\parallel}$. Here $\Phi_0$ is a constant that can be interpreted as the injection phase for an electron that starts its acceleration at $x \approx 0$. The only difference between the fields of the CP-LG and LP-LG beams is their amplitude. Therefore, we can skip the derivation here and directly apply the result of Ref.~\cite{shi2021a}. We have the following longitudinal momentum gain for an electron injected into the laser beam close to the peak of the envelope:
\begin{equation} \label{momentum gain}
    \frac{\Delta p_\parallel}{m_e c} = - a_* \frac{\pi^2 w_0^2}{\lambda_0^2} \left( \cos \Phi_0 - \cos \left[ \Phi_0  - 2\tan^{-1} (x/x_R) \right] \right), 
\end{equation}
where $a_*$ is the normalized amplitude of the longitudinal field. 

We obtain the terminal momentum gain by taking the limit of $x/x_R \rightarrow \infty$ in \cref{momentum gain}, which yields
\begin{equation}
    \frac{\Delta p_{\parallel}^{term}}{m_e c} = 2a_* \frac{\pi^2 w_0^2}{\lambda_0^2} \cos \left( \Phi_0 - \pi \right).
\end{equation}
One can understand the dependence on $\Phi_0$ by recalling that the electron is continuously slipping with respect to the forward-moving structure of $E_{\parallel}$ as it moves with the laser pulse. Delayed injection into the accelerating phase means that the electron slips into the decelerating phase before the amplitude of $E_{\parallel}$ becomes small due to the beam diffraction. As a result, the net momentum gain is reduced. The energy gain occurs only for $\pi/2 < \Phi_0 < 3 \pi /2$. The assumption that the electron is moving forward with ultra-relativistic velocity is no longer valid for $3 \pi / 2 < \Phi_0 < 5 \pi /2$, which, in term, invalidates the derived expression. It is useful to re-write our result in terms of electron energy. We assume that the electron experiences a considerable energy gain due to the increase of its longitudinal momentum, so that the terminal energy is $\varepsilon^{term} \approx c p_{\parallel}^{term} \approx c \Delta p_{\parallel}^{term}$. We then have
\begin{equation} 
    \frac{\varepsilon^{term}}{m_e c^2} = 2a_* \frac{\pi^2 w_0^2}{\lambda_0^2} \cos \left( \Phi_0 - \pi \right).
\end{equation}
We now take into account the expression for $a_*$ in terms of the period-averaged power $P$ given by \cref{a of P} to obtain the following practical expression:
\begin{equation} \label{max energy}
    \varepsilon^{term} [\mbox{GeV}] \approx 0.5 \cos (\Phi_0 - \pi) P^{1/2} [\mbox{PW}]. 
\end{equation}
In comparison to the acceleration by a CP-LG beam with the same power $P$~\cite{shi2021a}, the terminal energy in the LP-LG beam is lower by a factor of $\sqrt{2}$.

\begin{figure}
 \centering
 \includegraphics[width=0.45\linewidth]{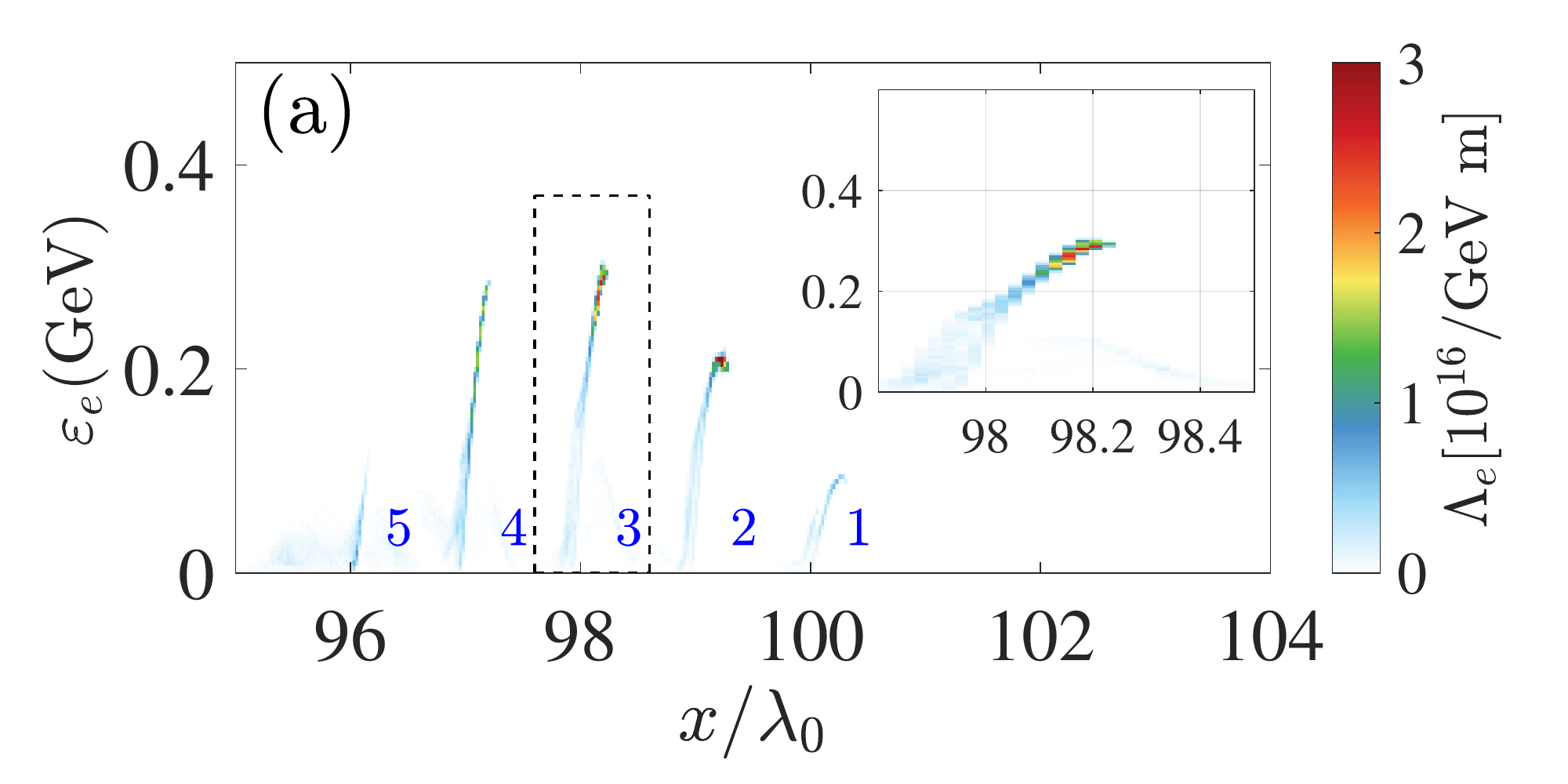} 
 \includegraphics[width=0.45\linewidth]{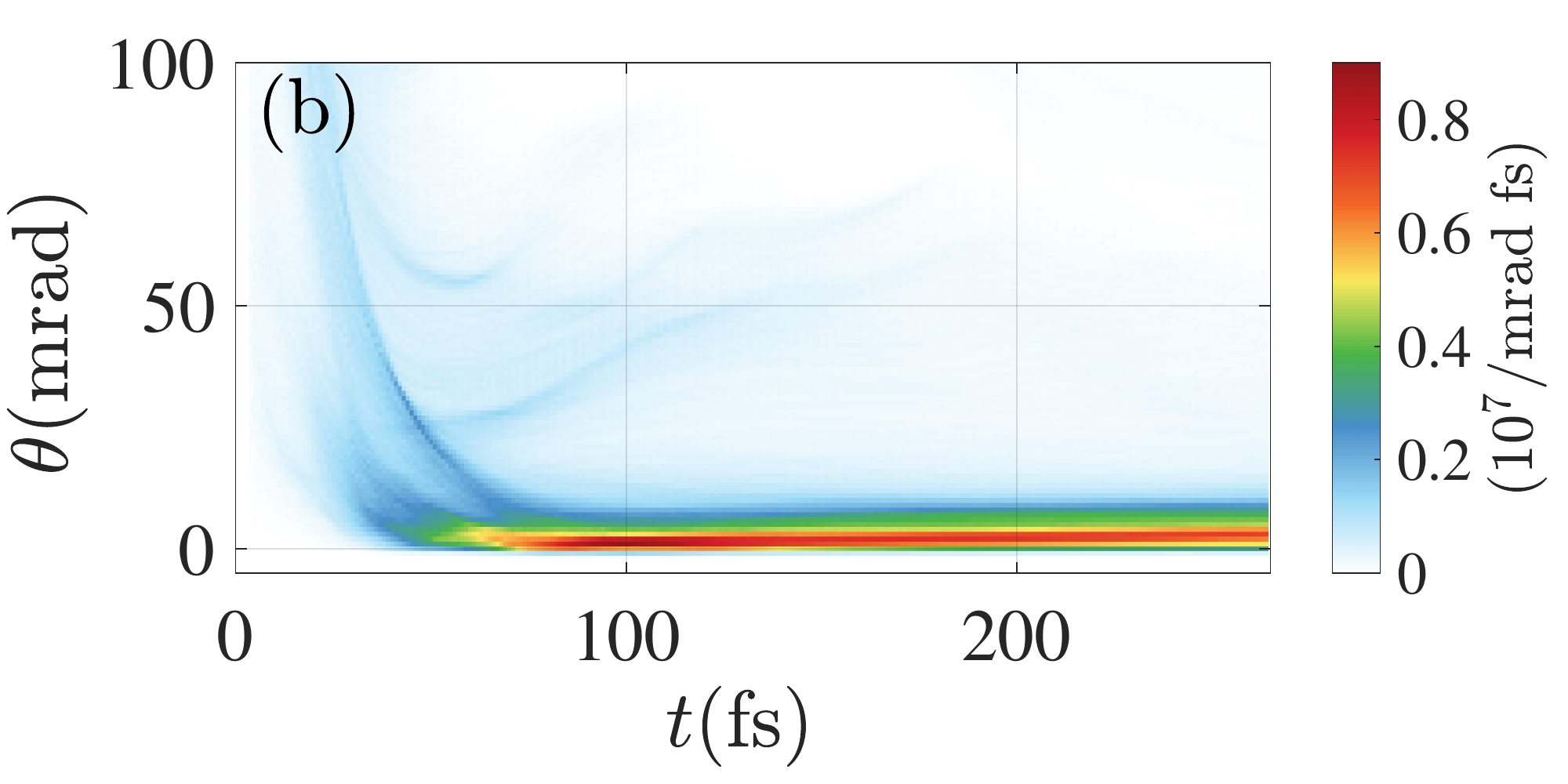} 
\includegraphics[width=0.45\linewidth]{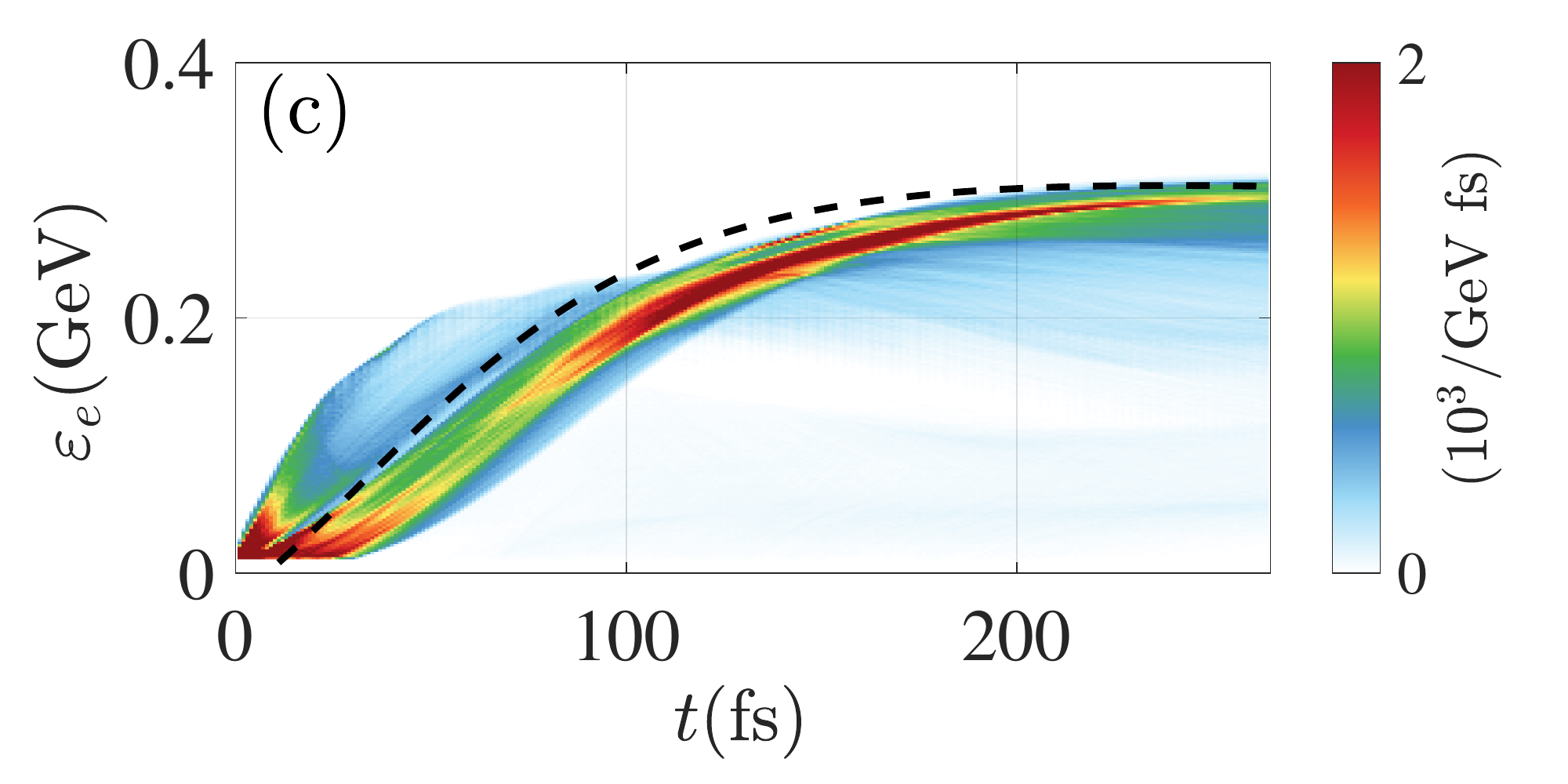} 
\includegraphics[width=0.45\linewidth]{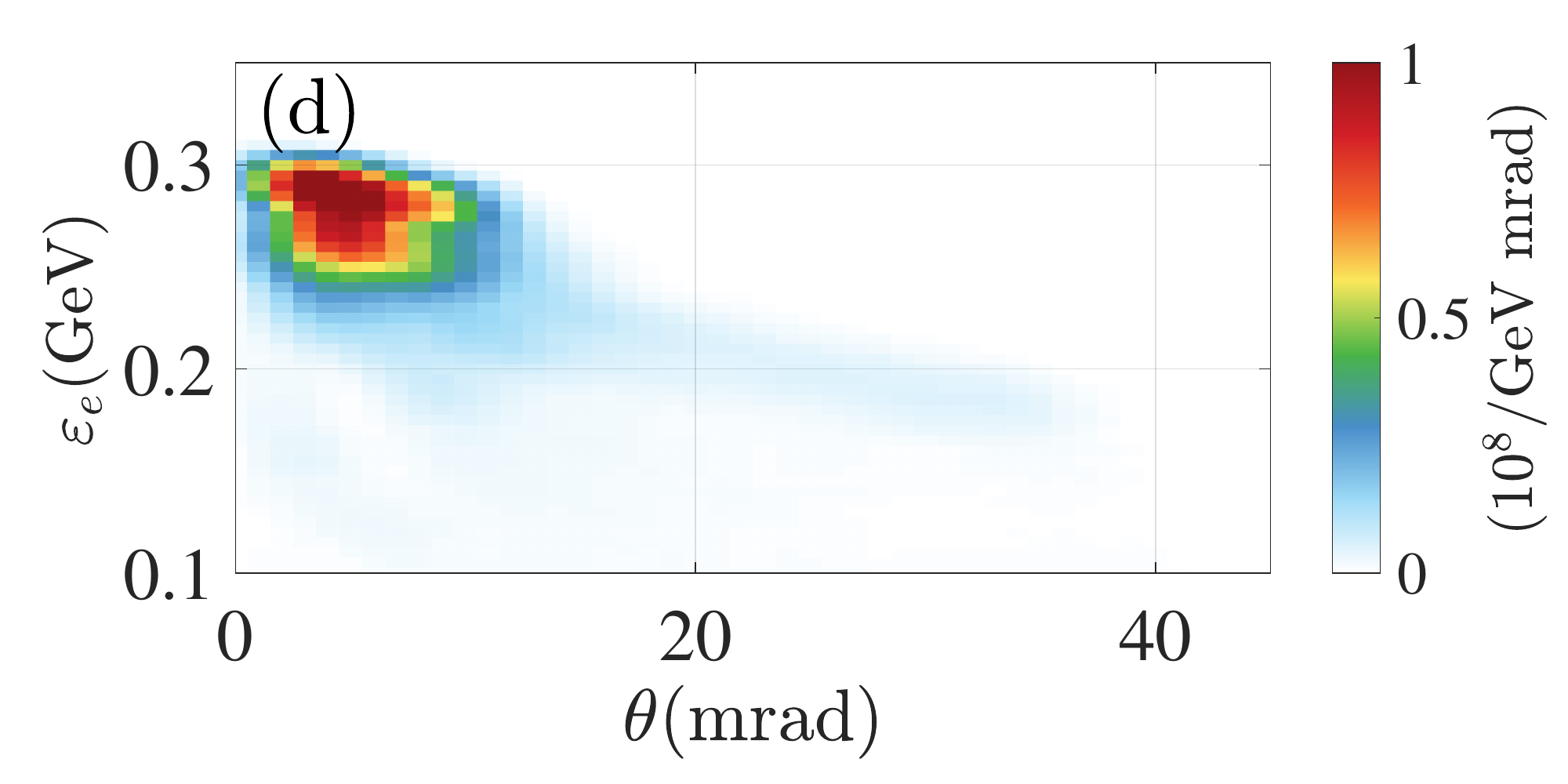}
\caption{Result of the long-term electron acceleration in the reflected LP-LG laser beam close to the beam axis. (a)~Electron energy distribution as a function of $x$ at $t = 261$~fs for electrons with $r<1.5~\micron$. 
The inset shows the third bunch that is marked with the dashed rectangle in the main plot. 
(b)~Time evolution of the electron distribution over the divergence angle $\theta$ in the third bunch ($r < 2 w_0$). (c)~Time evolution of the electron energy spectrum in the third bunch. The black dashed curve is the prediction obtained from \cref{momentum gain} with  $\Phi_0 = 0.8\pi$. The start time of the acceleration is used as an adjustable parameter. (d)~Electron energy versus the divergence angle in the third bunch shown in the inset of panel (a).} \label{ebunch261fs}
\end{figure}

\Cref{ebunch261fs} provides information of the long-term electron acceleration in the 3D PIC simulation. \Cref{ebunch261fs}(a) shows the electron energy distribution as a function of $x$ at $t = 261$~fs. {Note that the plot of $n_e$ integrated over the laser beam cross section is shown in \cref{Comp9fs261fs}(c) of \cref{append:conv}.} By this point, the electrons have roughly traveled a distance of $100 \lambda_0$ with the laser beam. Note that $t = 261$~fs is chosen as the time of the snapshot in order to facilitate a comparison with the results for the CP-LG beam presented in Ref.~\cite{shi2021a}. The pronounced bunching is maintained by the periodic accelerating structure of $E_{\parallel}$. The third bunch travels close to the peak of the laser envelope, which results in the highest electron energy gain. In what follows, we focus on this specific bunch. 

Figures~\ref{ebunch261fs}(b) and \ref{ebunch261fs}(c) show the time evolution of  the divergence angle and electron energy within the third bunch [see the dashed rectangle and the inset in \cref{ebunch261fs}(a)]. After an initial stage that lasts about 80~fs, the distribution over the divergence angle reaches its asymptotic shape. It can be seen from the snapshot in \cref{ebunch261fs}(b) [taken at $t = 261$~fs] that the bunch is monoenergetic, with most electron having a divergence angle that is less than 10~mrad. The time evolution of the energy spectrum, shown in \cref{ebunch261fs}(c), confirms that the bunch accelerates roughly as a whole. The dashed curve is the solution given by \cref{momentum gain}. We used the start time of the acceleration as an adjustable parameter because our model only captures the acceleration after the longitudinal motion becomes ultra-relativistic. The phase $\Phi_0$ is another adjustable parameter that we use to match the time evolution of the electron energy in the bunch. We find that $\Phi_0 \approx 0.8 \pi$ provides the best fit, whose result is shown in \cref{ebunch261fs}(c). The big energy spread at the early stage is likely due to the presence of the two lobes shown in \cref{edenst9fs}. The good agreement at later times indicates that our model captures relatively well the key aspects of the on-axis electron acceleration.

\begin{figure}
 \centering
 \includegraphics[width=0.75\linewidth]{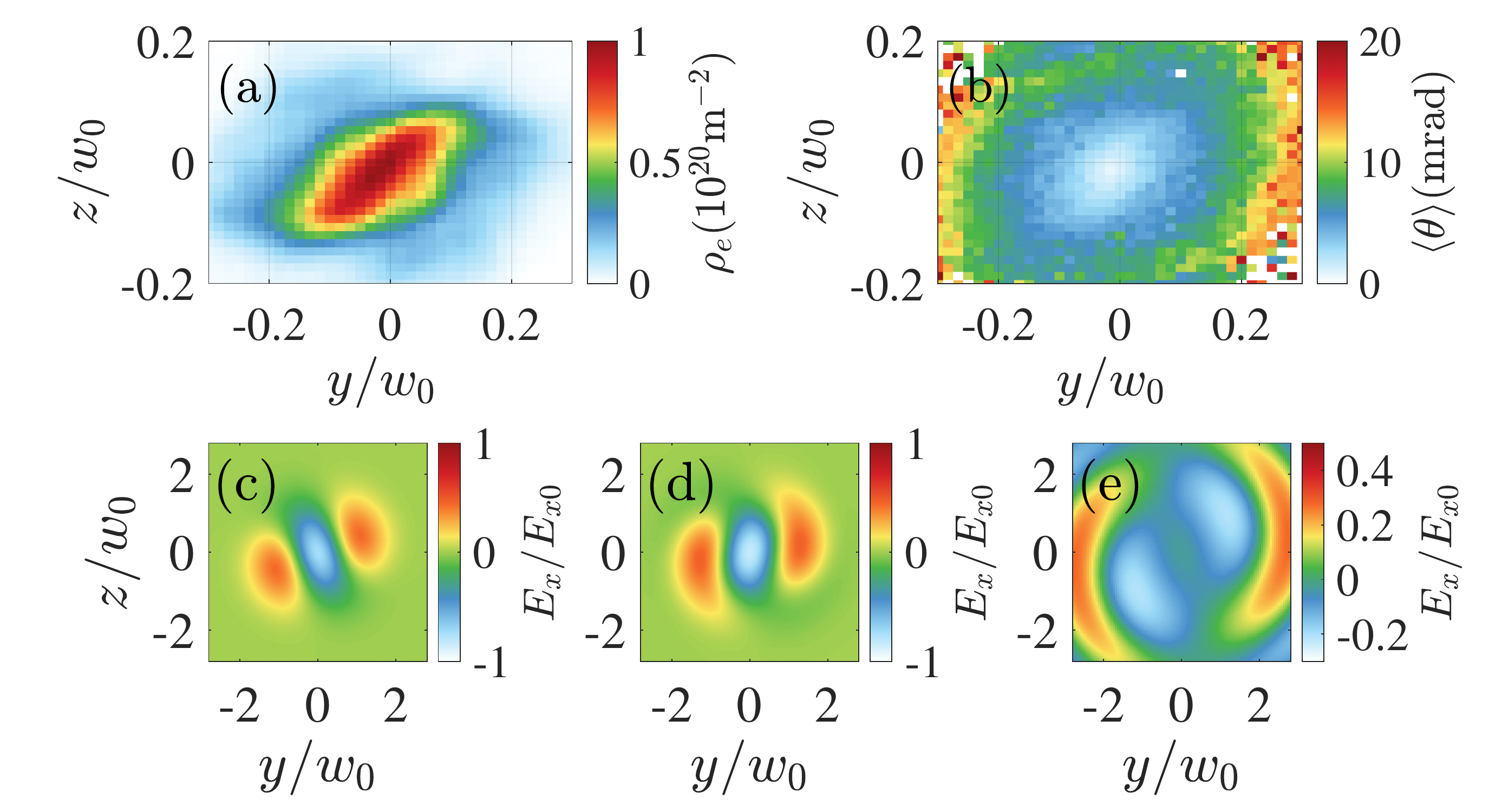} 
\caption{(a) areal density $\rho_e$ and (b) cell-averaged divergence angle $\theta$ in the cross-section of the third bunch at $t = 261$~fs and $\widetilde{x} = 2.3$.
(c), (d), and (e) Snapshots of the longitudinal electric field $E_x/E_{x0}$ in the cross section of the laser beam at: $\widetilde{x} = 0.1$, $t = 9$~fs (c),  $\widetilde{x} = 0.45$, $t = 46$~fs (d), and $\widetilde{x} = 2.3$, $t = 261$~fs (e). $E_x$ is calculated using the analytical expression~\cref{Ex_Tr} given in \cref{append:theory} and $E_{x0}$ is the amplitude of $E_x$ at $\widetilde{x} = 0$, $r = 0$.
} \label{ebunch261fs_tr}
\end{figure}

We find that the electron bunches retain noticeable asymmetry following their prolonged interaction with the laser beam. To illustrate the asymmetry, Figs.~\ref{ebunch261fs_tr}(a) and \ref{ebunch261fs_tr}(b) show the areal density $\rho_e$ and the cell-averaged divergence angle $\theta$ in the cross-section of the third bunch at $t = 261$~fs. The bunch asymmetry is likely imprinted by the asymmetry in $E_x$ shown in \cref{eb_xz}(e). Since the electron bunches are moving slower than the laser wavefronts, each bunch experiences a rotating $E_x$. This can be shown by examining the field structure at the location of a forward-moving ultra-relativistic electron bunch. We calculate $E_x$ in the beam cross section  using the analytical expression given in \cref{append:theory}. The longitudinal position is set by the expression $x = c t + \Phi_0\lambda_0/2\pi$ to mimic the longitudinal ultra-relativistic motion of an electron bunch. We set $\Phi_0 = 0.8\pi$, as this was the injection phase determined by our analysis.
Figures~\ref{ebunch261fs_tr}(c), \ref{ebunch261fs_tr}(d), and \ref{ebunch261fs_tr}(e)  show $E_x$ 
at $\widetilde{x} = 0.1$, 0.45, and 2.3. These locations correspond to the snapshots in Figs.~\ref{edenst9fs}, \ref{etraj}, and \ref{ebunch261fs}. The plots confirm that the field is indeed rotating, but they also show that the rotation is relatively slow, which is likely the reason why the asymmetry is retained by the electron bunch.

We conclude this section by providing additional parameters of the most energetic electron bunch (third bunch) generated by the considered 600~TW LP-LG laser beam. The electron energy in the bunch is 0.29~GeV with a FWHM of approximately $10\%$. The bunch has a charge of 9~pC and a duration of $\sim 270$~as. The divergence angle is as low as \dg{0.57} (10~mrad). The normalized emittance in $y$ is $\widetilde{\epsilon}_{rms, y} \approx 1.6 \times 10^{-6}$ and the normalized emittance in $z$ is $\widetilde{\epsilon}_{rms, z} \approx 1.4 \times 10^{-6}$.

\section{Summary and discussion}\label{Sec-5}

Using 3D PIC simulations, we have examined electron acceleration by a 600~TW LP-LG laser beam with $l = -1$ reflected off a plasma with a sharp density gradient. The simulations show that electrons can be effectively injected into the laser beam during its reflection. The electrons that are injected close to the laser axis experience a prolonged longitudinal acceleration by the longitudinal laser electric field. The simulations also show that the laser beam generates a train of mono-energetic ultra-relativistic electron bunches with a small divergence angle. \bc{The distinctive features of this acceleration mechanism are the formation of multiple sub-$\micron$ electron bunches, their relatively short acceleration distance (around $100~\micron$), and their high density (in the range of the critical density).}

An important conclusion from our study is that the key features that were previously reported for a CP-LG beam~\cite{Shi2021, shi2021a} are retained in the case of an LP-LG laser beam. It is likely that experimentally it will be easier to generate a high-power LP-LG beam than a high-power CP-LG beams. This is because the laser beams at high-power laser facilities are linearly polarized. Changing the polarization introduces additional challenges and complications that our approach of using an LP-LG beam allows to circumvent. We hope that this aspect will make it easier to perform a proof-of-principle experiment.

Even though there are key similarities, there are also differences in electron injection and acceleration between the cases of LP-LG and CP-LG beams. The injection into the LP-LG beam is more complex, leading to a formation of two side lobes that accompany the on-axis bunch. The asymmetry of the longitudinal electric field causes the on-axis electron bunches to become asymmetric. In contrast to that, the bunches generated by a CP-LG beam are axisymmetrical. For two beams with the same power, the LP-LG beam has weaker on-axis electric and magnetic fields. The reduction in the field strength leads to a reduced energy gain, with the terminal electron energy being lower by roughly a factor of $\sqrt{2}$. 

{Our mechanism relies on electrons becoming relativistic during the injection process. It is this feature that allows the injected electrons to surf with the laser pulse without quickly slipping from an accelerating phase into an adjacent decelerating phase. Since the longitudinal laser electric field plays a critical role in the injection process, its amplitude needs to be relativistic to generate relativistic injected electrons. A reduction of the incident laser power can thus degrade the mono-energetic spectra of the electron bunches by reducing the amplitude of the longitudinal field. To examine this aspect, we ran an additional simulation with a reduced incident power of 60~TW. Even though the laser still generates electron bunches in this case, the bunches are no longer mono-energetic. The peak energy is also noticeably lower than the value predicted by \cref{max energy}. The underlying cause is most likely the inability of electrons to stay for a prolonged period of time in an accelerating phase.}

{In this work, we primarily focused on the most energetic (third) electron bunch. The considered laser pulse generates five distinct electron bunches. Their parameters are given in \cref{table:train} of \cref{appendix: all bunches}. We want to point out that the front and tail of the considered laser pulse are steeper than what one would expect for a Gaussian pulse with the same FWHM, which was a deliberate choice made to reduce the size of the moving window and thus computational costs. 
The electrons must be relativistic during their injection, so that they can start moving with the laser beam without significant slipping. If this is not the case, then the mono-energetic feature discussed earlier might be hard to achieve. A dedicated study is required to determine the role of the temporal shape of the laser pulse and its overall duration. We anticipate that a longer laser pulse would produce a large number of ultra-relativistic electron bunches. For example, an 800~fs 600~TW LP-LG laser beam~\cite{zhu2018hpl} contains roughly 300 cycles, so it has the potential to generate a similar number of bunches. Such a pre-modulated electron beam with high charge can potentially be used to generate coherent undulator radiation and to create a free electron laser~\cite{Huang2012, MacArthur2018}. }

\section*{Data availability}

    The datasets generated and analyzed during the current study are available from the corresponding author on reasonable request.
    
\section*{Code availability}

    PIC simulations were performed with the fully relativistic open-access 3D PIC code EPOCH~\cite{EpochGit}.

\section*{\sffamily{Acknowledgements}}
Y S acknowledges the support by USTC Research Funds of the Double First-Class Initiative (Grant No. YD2140002003), Strategic Priority Research Program of Chinese Academy of Sciences (Grant No. XDA25010200) and Newton International Fellows Alumni follow-on funding. {D R B and A A  acknowledge the support by the National Science Foundation (Grant No. PHY 1903098).} Simulations were performed with EPOCH (developed under UK EPSRC Grants EP/G054950/1, EP/G056803/1, EP/G055165/1 and EP/M022463/1). The simulations and numerical calculations in this paper have been done on the supercomputing system in the Supercomputing Center of University of Science and Technology of China. {This research used resources of the National Energy Research Scientific Computing Center (NERSC), a U.S. Department of Energy Office of Science User Facility located at Lawrence Berkeley National Laboratory, operated under Contract No. DE-AC02-05CH11231.}



\appendix

\begin{table}[htb]
    \centering
    \begin{tabular}{|| c | c | c | c | c ||}
    \hline
    Sim. No. &Cell size & \makecell*[c]{Cell number \\(window size is the same)} & \makecell*[c]{Macro-particles per cell\\e ($r < 2.5\mathrm{\upmu m}$),\\ e ($r > 2.5\mathrm{\upmu m}$), C$^{+6}$} & \makecell*[c]{Order of\\ EM-field solver} \\
      \hline
    \#1& 1/40 $\micron$ & $400\times800\times800$ &200, 36, 24  &2\\
    \#2&1/40 $\micron$& $400\times800\times800$&400, 72, 48   &4\\
    \#3&1/50 $\micron$& $500\times1000\times1000$&200, 36, 24   &4\\
    \#4&1/80 $\micron$& $800\times1600\times1600$&100, 18, 12   &4\\
    \hline
    \end{tabular}
    \caption{Parameters used for the four simulations depicted in~\cref{Comp9fs261fs}. }
    \label{tab:domain}
\end{table}

\setcounter{equation}{0}
\setcounter{figure}{0}
\setcounter{table}{0}
\renewcommand\theequation{\Alph{section}\arabic{equation}}
\renewcommand\thefigure{\Alph{section}\arabic{figure}} 
\renewcommand\thetable{\Alph{section}\arabic{table}}   

\section{\sffamily{Convergence test}} \label{append:conv}

It was shown in Ref.~\cite{Shi2021} that the characteristics of the accelerated electron population in the setup considered in this paper can be sensitive to resolution used by the 3D PIC simulation. We ran a series of 3D PIC simulations using a 600~TW LP-LG beam with $l = -1$ and $p = 0$ to identify the simulation parameters that provide convergent results. In our convergence test, we varied the cell size and the number of macro-particles per cell. All relevant simulation parameters chosen for the convergence test are given in \cref{tab:domain}. The simulations were performed using EPOCH~\cite{Arber2015, EpochGit}.

\begin{figure}[htb]
 \centering
 \includegraphics[width=0.9\linewidth]{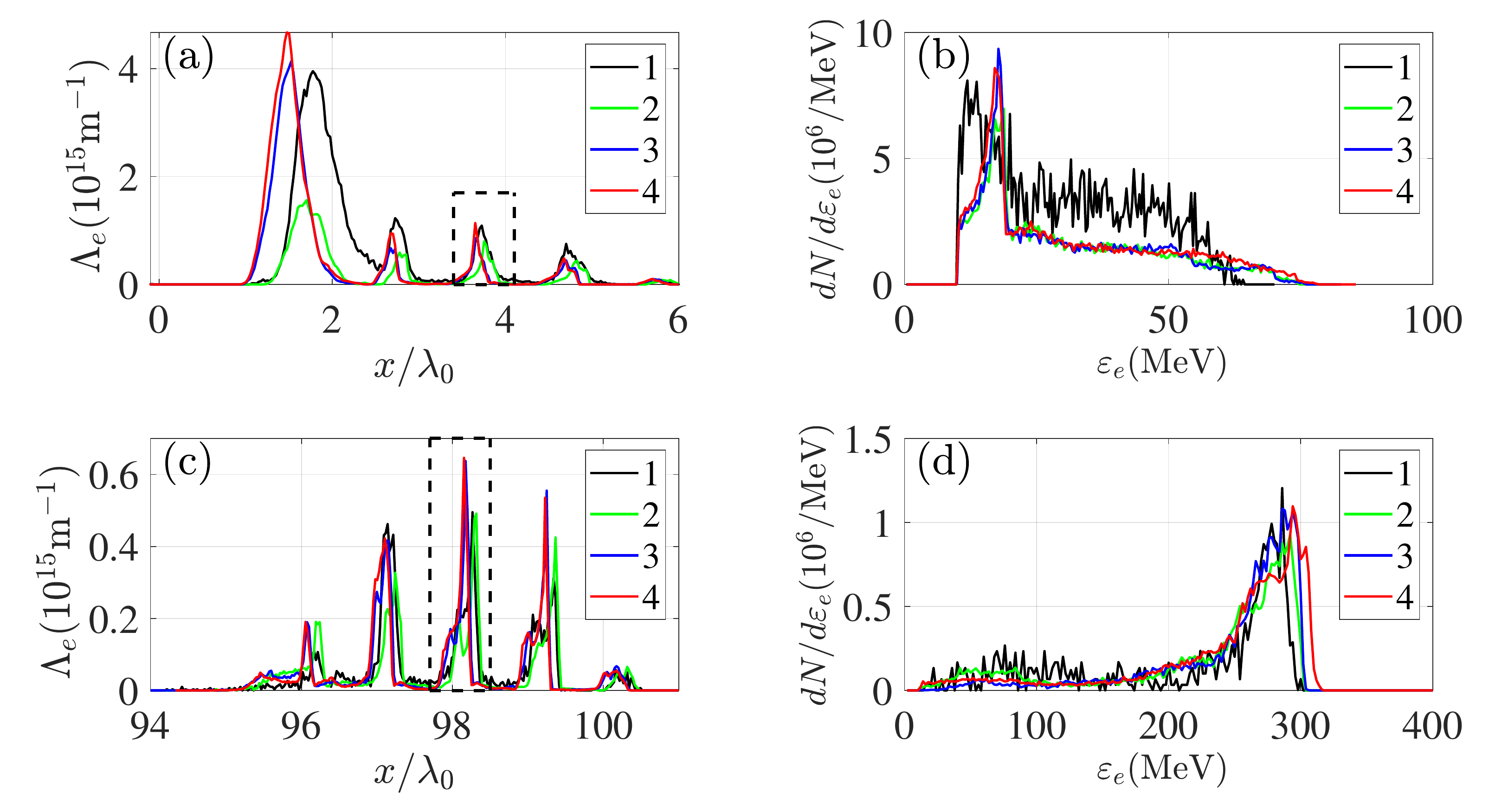}  
\caption{(a)\&(c) Linear density profiles early in the formation process of electron bunches, $t = 9$~fs [panel (a)], and after the bunches have experienced prolonged acceleration, $t = 261$~fs [panel (b)]. (c) Energy spectra of the third bunch (inside the dashed rectangle) from panel (a). (d) Energy spectra of the third bunch (inside the dashed rectangle) from panel (c). The legend in each plot provides the simulation number from \cref{tab:domain}. } \label{Comp9fs261fs}
\end{figure}

There are two features that we use to compare the simulations: the electron spectrum and the electron density. We use the linear density $\Lambda_e$, which is the number density integrated in the cross-section of the laser beam. \Cref{Comp9fs261fs}(b) shows $\Lambda_e$ early in the simulation at $t = 9$~fs. The formation of individual electron bunches is clearly visible in this plot. The curves for higher resolution simulations, i.e. simulations \#3 and \#4, are similar, which suggests that reducing the cell size below $1/80~\micron$ may not be necessary. \Cref{Comp9fs261fs}(d) shows four bunches generated by the laser pulse after they travel a significant distance in vacuum. Again, the curves for simulations \#3 and \#4 are very similar. 

Electron bunches experience longitudinal acceleration while they travel in vacuum with the laser pulse. It is therefore important to check not only the density of the bunches, but also their energy spectrum $dN /d\epsilon_e$, where $N$ and $\epsilon_e$ are the electron number and energy. \Cref{Comp9fs261fs}(b) shows the spectrum of the third bunch [inside the dashed rectangle] in \cref{Comp9fs261fs}(a). \Cref{Comp9fs261fs}(d) shows the spectrum of the same bunch [inside the dashed rectangle in \cref{Comp9fs261fs}(c)] after it has experienced extended acceleration. These spectra confirm that simulations \#3 and \#4 produce similar results. In the main text, we use the results of simulation \#4. This simulation is deemed to be reliable based on the presented convergence test.

{\section{Parameters of all the bunches generated by the LP-LG beam}} \label{appendix: all bunches}
 
\noindent {In the main text, we focused on the most energetic bunch, which is the third bunch out of the five bunches generated by the considered 600~TW LP-LG laser beam. \Cref{table:train} provides various parameters for all the five bunches. The parameters are calculated at $t = 261$~fs. }

\begin{table}[h]
\begin{center}
\begin{tabular}{|m{1.8cm} | m{1.7cm} | m{1.7cm} |m{2cm} |m{2cm} |m{1.5cm}|}
\hline
   & \#1 & \#2  & \#3 & \#4 & \#5\\
\hline
$\varepsilon_{e}$[GeV] ($\Delta \varepsilon_{e}/\varepsilon_{e}$) & 0.02$\sim$0.1 &0.02$\sim$0.28& 0.29 (10\%)& 0.22(6\%)& 0.1(15\%)\\
\hline
$\widetilde{\epsilon}_{rms, yz} [\mu$m]&0.95 & 0.88 & 1.5 & 0.64 & 0.92\\ 
\hline
$W$[mJ]&0.06 & 1.5& 2.2 & 1.3 & 0.06 \\ 
\hline
$Q$[pC]& 1.4&8& 9& 6.8& 0.7\\ 
\hline  
$\Delta t$[as]& 300 &360& 270& 260& 540\\ 
\hline 
\end{tabular}
\end{center}
\caption{\label{table:train}Parameters of all five electron bunches at $t$ = 261~fs. }
\end{table}

\section{Longitudinal electric field of an LP-LG laser beam} \label{append:theory}

Here we follow the notations introduced in Ref.~\cite{shi2021a} to provide analytical expressions for the fields of a linearly polarized helical beam in the paraxial approximation. Specifically, it is assumed that the diffraction angle, defined as $\theta_d = w_0/x_R$, is small, where $w_0$ is the beam waist, and $x_R = \pi w_0^2 / \lambda_0$ is the Rayleigh range. We consider a beam propagating along the $x$-axis. Its transverse electric field is polarized along the $y$-axis. It is convenient to normalize $x$ to $x_R$ and $y$ and $z$ to $w_0$ or $w(x)$:
\begin{eqnarray}
  && \widetilde{x} = x / x_R, \\
  && \widetilde{y} = y / w_0, \\
  && \widetilde{z} = z / w_0, \\
  && \widetilde{r} = \sqrt{ \widetilde{y}^2 + \widetilde{z}^2}.
\end{eqnarray}
The solution of the wave equation corresponding to an LG beam is given by
\begin{equation}\label{E_y_1}
    E_y = E_{0}  g(\xi) \exp(i \xi) \psi_{p,l} (\widetilde{x},\widetilde{r},\phi) 
\end{equation}
where $g$ is the envelope function with $\max (g) = 1$, 
\begin{equation}
    \xi \equiv 2\widetilde{x}/\theta_d^2 - \omega t
\end{equation}
is the phase variable,  and   
\begin{equation} \label{eq:psi}
    \psi_{p,l} (\widetilde{x},\widetilde{r},\phi) = C_{p,l} f(\widetilde{x})^{|l| + 1 + 2p} (1+\widetilde{x}^2)^p  L_{p}^{|l|}\left(\frac{2\widetilde{r}^2}{1+\widetilde{x}^2}\right) \left(\sqrt{2}\widetilde{r}\right)^{|l|} \exp \left[ -\widetilde{r}^2 f(\widetilde{x})\right] \exp\left(i l \phi\right) 
\end{equation}
is a mode with a  radial index $p$ and twist index $l$. Here we introduced
\begin{eqnarray}
  && \phi = \arctan \left( \widetilde{z} / \widetilde{y} \right), \\
  && f(\widetilde{x}) = \frac{1-i\widetilde{x}}{1+\widetilde{x}^2} = \frac{1}{\sqrt{1+\widetilde{x}^2}} \exp \left( - i \tan^{-1} \widetilde{x} \right).
\end{eqnarray}
The $L_{p}^{|l|}$ function is the generalized Laguerre polynomial and $C_{p, l}$ is a normalization constant. The modes $\psi_{p,l}(\widetilde{x},\widetilde{r},\phi)$ are orthonormal at a given  $\widetilde{x}$~\cite{Allen1992}, with
\begin{equation}
    C_{p,l} = \sqrt{\frac{2p!}{\pi (p + |l|)!}},
\end{equation}
such that
\begin{equation} \label{eq:norm}
    \int_0^{2\pi} d \phi \int_{0}^{\infty} \psi_{l,p}(\widetilde{x},\widetilde{r},\phi) \psi^*_{p,l}(\widetilde{x},\widetilde{r},\phi) \widetilde{r} d \widetilde{r} = 1.
\end{equation}
The period-averaged power in this beam is
\begin{equation} \label{P}
    P = \frac{c w_0^2}{8 \pi} E_0^2,
\end{equation}
where $c$ is the speed of light. Note that $E_0$ is not the peak amplitude of $E_y$ in the case of an LG beam.

The mode considered in the main text has $p = 0$ and $l = -1$. It then follows from \cref{E_y_1} that
\begin{equation}
    E_y = E_{0} g(\xi) C_{0, -1} [f(\widetilde{x})]^2 \sqrt{2}\widetilde{r} \exp \left[ -\widetilde{r}^2 f(\widetilde{x})\right] \exp\left(-i \phi\right) \exp( 2i\widetilde{x}/\theta_d^2 - i\omega t).
\end{equation}
As shown in Ref.~\cite{shi2021a}, the corresponding longitudinal electric field is 
\begin{equation} \label{E_x4ey}
  E_x = \frac{i \theta_d}{2} \left[ \frac{1}{\widetilde{r}} e^{ i \phi} - 2f\widetilde{r} \cos \phi \right] E_y,
\end{equation}
where it is taken into account that $p = 0$ and $l = -1$. The longitudinal electric field on the axis is given by
\begin{eqnarray}
  && E_{\parallel} \equiv E_x (\widetilde{r} = 0) = 
 \frac{i \theta_d}{\sqrt{\pi}} \frac{E_{0} g(\xi)}{{1+\widetilde{x}^2}} \exp( 2i\widetilde{x}/\theta_d^2 - 2 i \tan^{-1} \widetilde{x} - i\omega t).
\end{eqnarray}
It is convenient to re-write this expression by introducing phase
\begin{equation}
    \Phi = 2\widetilde{x}/\theta_d^2 - 2 \tan^{-1} \widetilde{x} - \omega t
\end{equation}
and amplitude
\begin{equation}
    E_* \equiv \theta_d E_0 / \sqrt{\pi},
\end{equation}
so that
\begin{equation} \label{E longitudinal}
   E_{\parallel}  =
  \frac{i E_* g(\xi)}{{1+\widetilde{x}^2}} \exp(i \Phi).
\end{equation}
The expression for $P$, recast in terms of the normalized amplitude
\begin{equation}
    a_* \equiv \frac{|e| E_*}{m_e c \omega},
\end{equation}
reads
\begin{equation}
    P = a_*^2 \frac{\pi^4}{2} \left( \frac{w_0}{\lambda_0} \right)^4 \frac{m_e^2 c^5}{e^2} .
\end{equation}
It follows from this relation that
\begin{equation} \label{a of P}
    a_* \approx 50 \left( \frac{\lambda_0}{w_0} \right)^2 \left(P \mbox{ [PW]} \right)^{1/2}.
\end{equation}


In the main text, we examine the field structure of $E_x$ away from the axis. The corresponding expression follows from \cref{E_x4ey}:
\begin{eqnarray}
  E_x = i E_{x0} e^{i\Phi_{*}}  \frac{e^{-r_{*}^2}}{1 + \widetilde{x}^2}  \left[ 1  - 2(1-i\widetilde{x}) r_{*}^2 \cos \phi e^{ - i \phi} \right] ,
\end{eqnarray}
where, for compactness, we used the following notations:
\begin{eqnarray}
  && y_{*} \equiv y \left/ w_0 \sqrt{1 + \widetilde{x}^2} \right. ,\\
  && z_{*} \equiv z \left/ w_0 \sqrt{1 + \widetilde{x}^2} \right. ,\\
 && r_{*} \equiv \widetilde{r} \left/ \sqrt{1 + \widetilde{x}^2} \right. ,\\
 && \Phi_{*}  \equiv  2\widetilde{x} \left/ \theta_d^2 \right.  + \widetilde{x} r_{*}^2 - 2 \tan^{-1} (\widetilde{x}) - \omega t, \\
 && E_{x0} \equiv (C_{0, -1}/\sqrt{2}) \theta_d E_{0} g(\xi).
\end{eqnarray}
The real part, under the assumption that $g(\xi)$ has no imaginary part, is given by
\begin{equation} \label{Ex_Tr}
  \mbox{Re}(E_x) = - E_{x0}\frac{e^{-r_{*}^2}}{1 + \widetilde{x}^2}  \left[ (1 - 2 y_*^2 + 2 \widetilde{x} y_* z_*) \sin \Phi_* + 2y_*(\widetilde{x} y_* +  z_*) \cos \Phi_* \right],
\end{equation}
where $y_* = r_* \cos \phi$ and $z_* = r_* \sin \phi$.


\bibliographystyle{ieeetr}

\end{document}